\documentclass[journal]{IEEEtran}

\usepackage[T1]{fontenc}
\usepackage{mathtools}
\usepackage{xurl}
\usepackage{amsmath}
\usepackage{amssymb}
\usepackage{amsthm}
\usepackage{amsfonts}
\usepackage{upgreek}
\usepackage{dirtytalk}
\usepackage{optidef}
\usepackage{subcaption}
\usepackage{algorithmic}
\usepackage{algorithm}
\usepackage{array}
\usepackage{textcomp}
\usepackage{stfloats}
\usepackage{url}
\usepackage{verbatim}
\usepackage{graphicx}
\usepackage{setspace}
\usepackage{calc}
\usepackage{enumerate}
\usepackage[usenames]{color}
\usepackage{upref}
\usepackage{times}
\usepackage{hyperref}
\usepackage{xcolor}
\usepackage{cite}
\usepackage{pgfplots}
\pgfplotsset{compat=1.18}

\begin{document}

\title{Q-Backbone: A Quantum-Enhanced Control Plane for Future Communication Networks}
\author{\IEEEauthorblockN{Mahdi Chehimi, Nour Dehaini, Nikos A. Mitsiou,  Ioannis Krikidis, Gan Zheng}
\thanks{Mahdi Chehimi and Nour Dehaini are with the Department of Electrical and Computer Engineering, American University of Beirut, Beirut, 1107 2020, Lebanon. Emails: \{mc127,nad29\}@aub.edu.lb.

Nikos A. Mitsiou and Ioannis Krikidis are with the Department of Electrical and Computer Engineering, University of Cyprus, Nicosia, Cyprus. Emails: \{nmitsi02,krikidis\}@ucy.ac.cy.

Gan Zheng is with the School of Engineering, University of Warwick,
Coventry, CV4 7AL, UK. Email: gan.zheng@warwick.ac.uk.}
\vspace{-.2cm}}
\maketitle

\begin{abstract}
Future networks will need to make network-wide decisions, including traffic engineering, network slicing, and wireless optimization, under strict latency, energy, and reliability constraints. The computational complexity of these problems increasingly challenges classical optimization methods. This article proposes Q-Backbone~(QB), a quantum-enhanced control plane for communication networks in which quantum processing units~(QPUs) operate alongside classical computing resources as accelerators for network intelligence. QB is designed as a four-layer architecture that combines heterogeneous infrastructure, hybrid quantum-classical runtime services, policy-driven task orchestration, and communication-network applications. A central component of QB is the Quantum Invocation Policy~(QIP), which dynamically determines when quantum acceleration is beneficial and when classical execution should be preferred. A case study on deadline-aware orchestration of distributed quantum jobs over heterogeneous QPUs shows that QB can improve workload execution under tight deadline constraints, serving up to 25\% more jobs than existing quantum-cloud scheduling baselines. Finally, open challenges and opportunities towards the deployment of QB are highlighted and discussed.
\end{abstract}
\begin{IEEEkeywords}
Quantum-enhanced networking, distributed quantum computing, quantum optimization, quantum machine learning.
\end{IEEEkeywords}
\IEEEpeerreviewmaketitle

\section{Introduction}\label{sec:intro}

\IEEEPARstart{T}{he} transition from 5G to 6G and beyond demands that communication networks evolve into intelligent computing fabrics rather than simple transport infrastructures. Joint decisions across backbone control, physical layer optimization, and machine learning (ML)-driven network management can involve up to thousands of coupled binary variables, competing objectives, and strict control-loop deadlines, creating combinatorial workloads that increasingly challenge classical optimization at scale~\cite{9390169}. Although classical ML methods can improve prediction and policy quality, the underlying decision stage still relies on classical computation, so the combinatorial bottleneck persists. Quantum-enhanced acceleration is positioned precisely to address this remaining gap through hybrid quantum-classical computing architectures.
Recent advances have sparked interest in using hybrid quantum-classical approaches for communication and networking related workloads, demonstrating the potential benefits of quantum-enhanced optimization and learning techniques for network control~\cite{fan2022hybrid}. In parallel, ongoing progress in distributed quantum computing~(DQC)~\cite{barral2025dqc} and quantum communication~\cite{chehimi2022physics} is gradually enabling the execution of these algorithms using multiple quantum processing units (QPUs). However, several key questions remain open before quantum resources can be integrated into future communication networks. In particular, the questions of \emph{where} QPUs should be deployed, \emph{which} and \emph{when} workloads should invoke a potential quantum acceleration, and \emph{how} quantum and classical resources should be jointly coordinated remain unexplored.


This article aims to answer these questions, noting that the metropolitan and core network points of presence~(PoPs), that comprise the \textbf{backbone} infrastructure, together with regional data centers, should be the primary deployment sites for QPUs. These sites already host dense computing power, including software-defined networking (SDN) controllers, and high performance computing clusters. In addition, backbone PoPs already aggregate network telemetry, routing, and orchestration while operating centralized control loops on timescales compatible with current QPU overheads. This is in contrast to the resource-constrained edge, where QPU deployment is impractical today.  Building on this observation, the article proposes \emph{Q-Backbone~(QB)}, a quantum-enhanced control plane for future communication networks. 

QB follows a four-layer architecture. Layer~1 comprises the distributed computing infrastructure, where central processing units (CPUs), graphics processing units (GPUs) and QPUs are deployed across backbone PoPs and regional data centers.  Layer~2 provides the hybrid runtime environment responsible for preparing, executing, and validating quantum-classical workloads. Layer~3 implements the orchestration and control logic which determines how workloads are mapped across available resources.  A central component of this layer is the Quantum Invocation Policy~(QIP), which determines, according to the problems' characteristics, operational context, hardware state of the network and the capabilities of single QPUs,   whether a workload should be executed through quantum or classical resources.  Layer~4 exposes hybrid quantum-classical optimization and learning workloads to the lower layers, including, but not limited to, backbone control, and wireless physical-layer applications. Finally, a first evaluation of the proposed Layer-3 scheduler is presented under realistic near-term quantum hardware constraints and tight control-loop deadlines. Specifically, the study considers the deadline-aware orchestration of concurrent quantum workloads whose computational requirements exceed the capacity of individual QPUs, thereby requiring distributed execution across multiple quantum resources.
\vspace{-0cm}
\section{Q-Backbone: Architecture and Operation}\label{sec:architecture}

The value of a quantum-enhanced computing plane for 6G backbones depends not only on its architectural structure, but also on how it operates at runtime. For this reason, QB's architecture and runtime operation are presented as a unified design. In QB, the deployment environment shapes the layered architecture, the layered architecture determines the orchestration capabilities available at runtime, and the runtime policies ultimately decide when quantum resources should be invoked. Accordingly, QB combines three main elements, a backbone-centric deployment model, an SDN/NFV-compatible four-layer architecture that integrates QPUs alongside classical accelerators, and QIP that determines whether workloads should execute through quantum or classical resources.

\begin{figure*}[t]
\centering
\includegraphics[width=\textwidth]{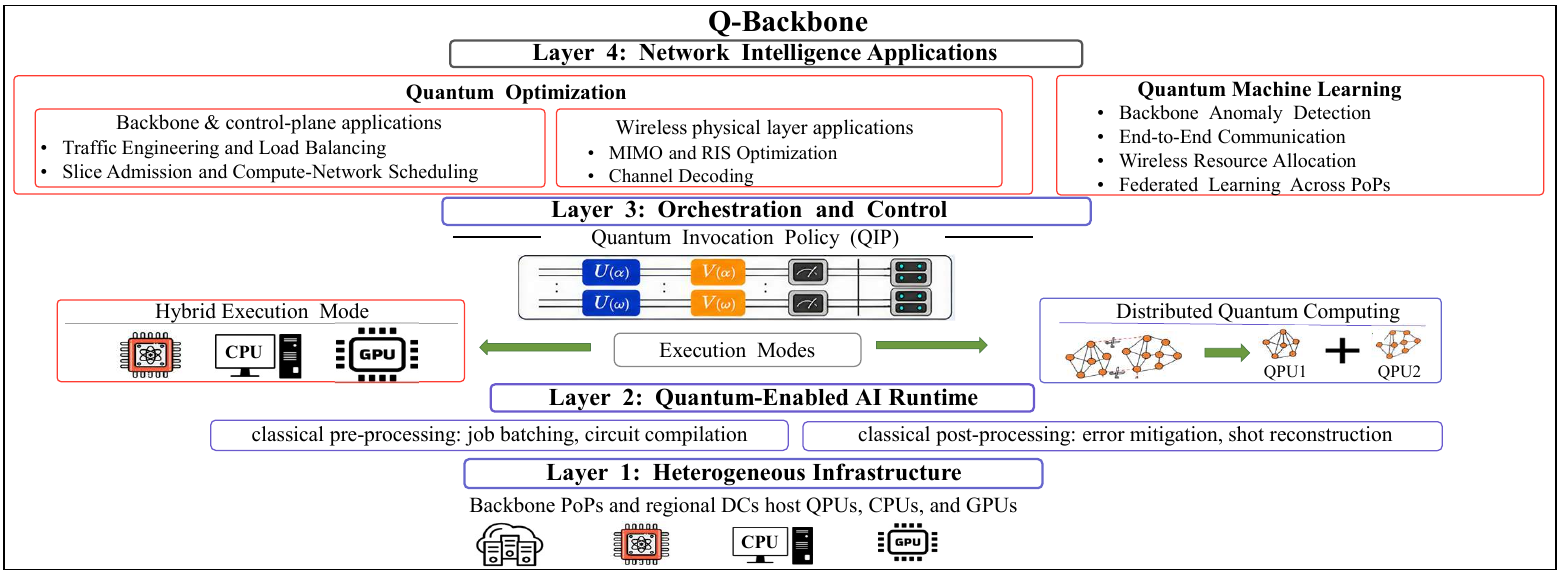}
\caption{Q-Backbone four-layer architecture. Layer~1 hosts heterogeneous CPU/GPU/QPU infrastructure distributed across backbone PoPs and regional DCs. Layer~2 provides hybrid quantum-classical runtime services, classical pre/post-processing, circuit compilation, error mitigation, job batching, and shot reconstruction. Layer~3 implements the policy-driven orchestration: the QIP routes each admitted job to single-site hybrid execution or DQC. Layer~4 exposes quantum optimization and QML workloads, spanning from backbone and control-plane applications to wireless physical-layer problems.}
\label{fig:QB_architecture}
\end{figure*}

\subsection{Backbone as the Deployment Site}\label{sec:why_backbone}

QB adopts a \emph{co-located integration} model in which QPUs are physically hosted alongside GPU/CPU clusters in backbone PoPs and regional DCs, interconnected via a high-speed internal fabric. Targeting the backbone rather than the edge as the practical near-term deployment point follows from the following mutually reinforcing factors. First, backbone sites already provision the power and cooling required for dense GPU clusters, making the cryogenic or vacuum-chamber infrastructure of leading QPU modalities feasible, a non-starter at edge nodes. Second, backbone controllers possess cross-domain visibility over traffic matrices, multi-PoP slice states, and inter-DC compute, so they aggregate precisely the inputs that large-scale optimization workloads consume. Moreover, the workloads themselves, i.e., the multi-commodity TE across dozens of PoPs, the joint slice admission over hundreds of requests, and the end-to-end resilience planning-are large enough to push classical solvers to their limits, yet structured enough to admit Ising encodings amenable to QPU acceleration. Finally, backbone control loops re-optimize on seconds-to-minutes timescales, comfortably accommodating QPU compile-execute-postprocess overheads that would be intolerable at sub-millisecond edge deadlines. This deployment posture is structural rather than incidental. It is actually what makes a quantum-enhanced computing plane technically relevant today, by aligning the strength of QPUs with the temporal and combinatorial regime of network control.

\subsection{Layered Architecture}\label{sec:layers}

QB augments existing SDN/NFV control planes with a quantum-aware compute substrate organized in four layers, as shown in Fig.~\ref{fig:QB_architecture}. To our knowledge, this is the first layered abstraction to treat the QPU as a first-class accelerator alongside CPUs and GPUs in a communication control plane, rather than as an external service called over an opaque API.

The \emph{infrastructure layer} (Layer~1) hosts CPUs, GPU clusters, and QPUs (gate-based or annealers) across backbone PoPs and regional DCs, with classical high-capacity optical transport interconnecting sites. Quantum interconnections for DQC are treated as a future option rather than a baseline assumption, keeping the architecture deployable on contemporary hardware while leaving room to absorb quantum networking when it matures \cite{caleffi2024dqc}.

The \emph{quantum-enabled AI runtime layer} (Layer~2) manages hybrid job lifecycles. It performs classical preprocessing, including problem reduction to quadratic unconstrained binary optimization (QUBO) form, coordinates QPU execution via quantum algorithms such as quantum approximate optimization algorithm (QAOA), variational quantum eigensolver (VQE) or Grover's algorithm, and applies classical post-processing such as solution repair and local search. It additionally provides circuit compilation, error mitigation, job batching, shot-level result reconstruction for cut circuits, and a  quantum machine learning (QML) sub-runtime that supports variational circuits and quantum kernel estimation under the same lifecycle envelope as optimization workloads.

The \emph{orchestration and control layer} (Layer~3) is where QB diverges most sharply from conventional network orchestrators. A quantum-aware scheduler maps each job to the best accelerator mix; a solver and model registry stores pre-validated QUBO templates, QAOA configurations, and QML checkpoints; a telemetry engine continuously feeds network state into applications; and the QIP, detailed in Section~\ref{sec:qip}, governs whether each job is routed to a QPU or to a classical fallback. When a job's reduced circuit exceeds single-QPU capacity, the same scheduler partitions subcircuits across QPUs, distributes their sampling shots, and respects any precedence constraints induced by circuit-cutting techniques. Conventional SDN and network function virtualization (NFV) orchestrators, designed around CPU/GPU heterogeneity, have no equivalent of quantum-eligibility assessment, network-aware quantum job routing, or QPU lifecycle management; Layer~3 supplies these missing capabilities.

The application layer (Layer~4) exposes accelerator-agnostic control primitives to closed-loop network controllers, including \texttt{optimize} for  network and physical-layer optimization tasks, \texttt{predict} for traffic and network-state forecasting related tasks, and \texttt{detect} for telemetry-driven anomaly and fault detection. These primitives support applications spanning traffic engineering, network slicing, and network monitoring. Note that applications remain unaware of whether execution used CPU, GPU, QPU, or a hybrid mix, freeing controller logic from accelerator-specific assumptions and allowing the QIP to redirect jobs as hardware availability and problem features change. Thus, this abstraction generalizes the manual, per-problem decomposition strategy of~\cite{fan2022hybrid}, where integer programming subproblems are routed to quantum solvers and continuous parts remain classical.

\subsection{Runtime Orchestration and the Quantum Invocation Policy}\label{sec:qip}

While the layered architecture defines what QB \emph{is}, its value emerges only at runtime, when applications submit jobs and Layer~3 must decide how to execute them. Two design choices distinguish QB's runtime from a conventional SDN/NFV runtime: a uniform \emph{reduce-solve-verify} pipeline that every job traverses, and a decision function-the QIP-that dynamically chooses whether to invoke a QPU at all.

\begin{figure}[t]
\centering
\includegraphics[width=0.95\columnwidth]{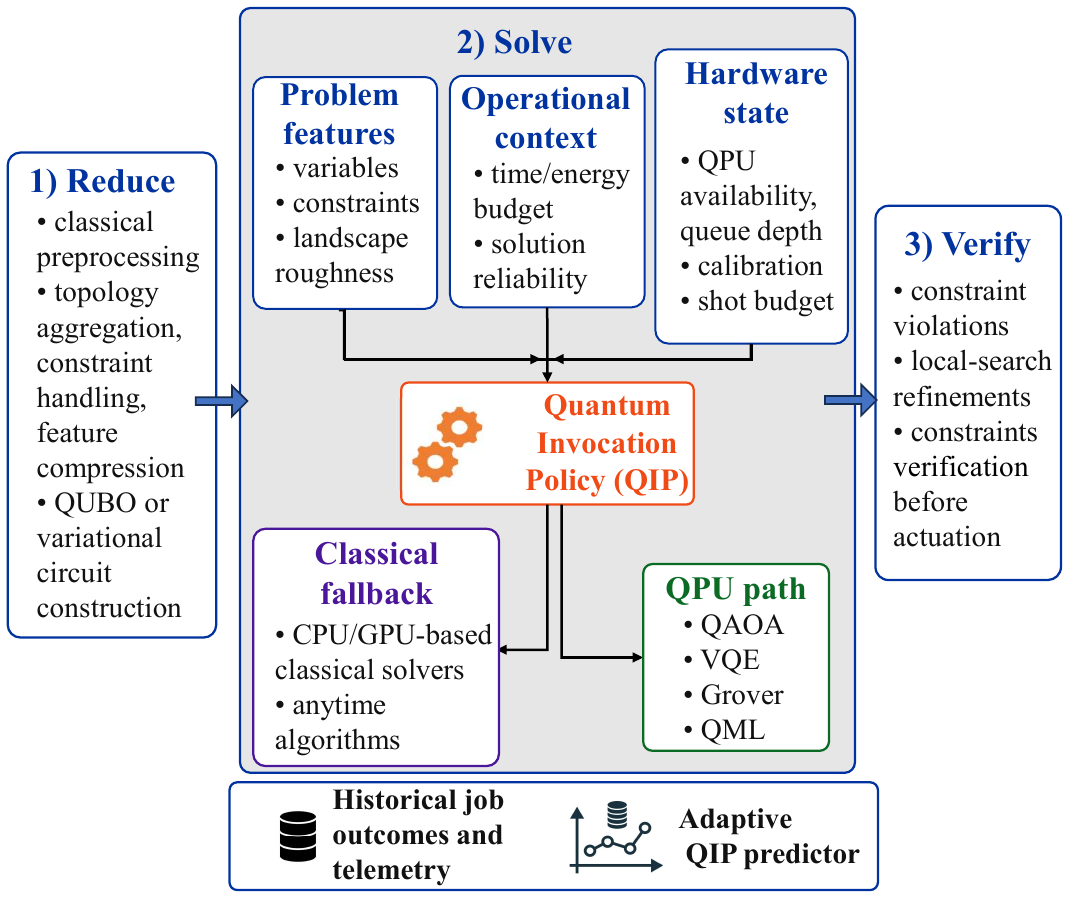}
\caption{The reduce–solve–verify runtime pipeline of QB. Jobs are first classically reduced to a tractable QUBO or variational circuit, then solved under the control of the QIP, and finally verified against actual operational constraints before actuation. }
\label{fig:QIP}
\vspace{-0cm}
\end{figure}

Every job submitted to QB follows a three-phase pipeline. In the \emph{reduce} phase, classical preprocessing coarsens the problem through topology aggregation, demand clustering, constraint screening, or feature compression, and constructs either a QUBO or a variational circuit of tractable size. In the \emph{solve} phase, the QIP routes the reduced problem to either a QPU or to a classical fallback. We note that when the QIP selects the quantum path, the Layer-3 scheduler realizes it either through single-site hybrid execution or, when the reduced circuit exceeds one QPU, through distributed execution as it wil be described in Section~II-D. In the \emph{verify} phase, classical post-processing repairs constraint violations, performs local-search refinement, and validates the solution against operational constraints before any actuation. The same pipeline applies uniformly to optimization and to QML inference, with the variational circuit replacing the QUBO step in the latter case-a uniformity that lets the orchestrator schedule heterogeneous workloads under a single invocation contract.


The novelty of the runtime layer lies in the QIP itself. Existing SDN/NFV orchestrators schedule workloads across CPUs and GPUs, but they do not decide whether a QPU should be invoked for a given control-loop task. The QIP fills this gap by evaluating each job along three dimensions. \emph{Problem features} capture the size and structure of the reduced problem, including its variable count, constraint density, and estimated landscape roughness. \emph{Operational context} captures the remaining control-loop deadline, reliability target, and energy budget, which determine whether a probabilistic quantum solution is acceptable. \emph{Hardware state} captures QPU availability, queue depth, calibration freshness, and feasible shot budget within the time window. If these inputs indicate that quantum execution is unlikely to improve the result before the deadline, the job is routed to a classical CPU/GPU fallback, such as a parallel heuristic or mixed-integer solver. Over time, the QIP can be refined using historical job outcomes and network telemetry, allowing QB to learn when quantum invocation is operationally worthwhile.

\vspace{-0.2cm}
\subsection{Execution Modes: Single-Site Hybrid and Distributed}\label{sec:exec_modes}

The reduce-solve-verify pipeline can run in either of two execution modes, selected jointly by the QIP and the Layer-3 scheduler.  The default and most practical mode in the noisy intermediate-scale quantum~(NISQ) era is \emph{single-site hybrid execution}, where the full workflow remains within one PoP that hosts CPUs, GPUs, and a local QPU. Classical resources perform reduction, orchestration, parameter updates, and verification. Then, if the QIP approves quantum acceleration, the solve step is offloaded to the local QPU, otherwise it is completed by the local classical fallback. This keeps both QPU-assisted and fallback execution within the same site, avoiding the coordination overhead of distributed execution.


\emph{Distributed quantum computing} is a secondary mode, invoked when three conditions hold simultaneously: the problem decomposes naturally into weakly coupled subproblems, no single QPU has sufficient capacity to absorb the reduced circuit, and the inter-site coordination overhead is small relative to the solve time. Spatial decomposition-partitioning the backbone by region, solving each region's sub-problem on a local QPU, and reconciling inter-region flows classically-is the most natural fit for backbone control workloads. Two DQC realizations are possible. \emph{Classical-coordination DQC}, in which subcircuits exchange only classical results across the backbone, is the only practical mode in NISQ-era deployments; QB exercises this mode in the case study of Section~\ref{sec:usecase}. \emph{Quantum-networked DQC}, in which entanglement is exchanged between QPUs over a quantum interconnect~\cite{chehimi2022physics}, remains a research frontier that the architecture is designed to absorb transparently when available. Recent experimental demonstrations combining multiple quantum processors with real-time classical communication~\cite{carrera2024combining} indicate that near-term DQC is becoming practically feasible, provided the control plane is aware of the local-operations-and-classical-communication (LOCC) induced precedence constraints that circuit cutting introduces, an awareness exercised by the Layer-3 scheduler in the case study to follow.


\section{Applications to Communications Networks}\label{sec:applications}

Communication networks give rise to a wide range of large-scale, delay-sensitive problems that are natural candidates for quantum-enhanced solvers. Layer 4 of the QB provides the architectural umbrella through which such problems, spanning backbone control to wireless physical layer design, can be formulated, decomposed, and supported in real time via hybrid and distributed quantum-classical execution. We organize these problems into two classes, combinatorial problems that can be cast as Ising optimization instances, and learning-based tasks for which QML can be implemented in the QB runtime.

\subsection{Quantum Optimization for Network and Wireless Control}\label{sec:qopt}

A representative gate-based model for this class of problems is the Quantum Approximate Optimization Algorithm~(QAOA). When the optimization objective is written in Ising form, linear and quadratic terms can be mapped directly into single- and multi-qubit quantum operations, making QAOA particularly attractive for combinatorial communication and networking problems ~\cite{Krikidisris}.

Some representative examples follow.
\subsubsection{Backbone and control-plane applications}\label{sec:backbone_apps}
\begin{itemize}
\item \emph{Traffic engineering and load balancing}: Traffic engineering in a capacitated backbone can be formulated as a multi-objective routing problem in which metrics such as delay, jitter, transmission loss, and throughput must be jointly optimized. In this setting, routing objectives and feasibility constraints can be encoded as quadratic cost and penalty terms, yielding a QAOA formulation.

\item \emph{Slice admission and compute-network scheduling}: Joint slice admission, VNF placement, and path selection form a coupled NP-hard optimization problem. Following the reduce-then-solve pattern, the original mixed continuous-binary formulation can be decomposed into a discrete and a classical subproblem, with the placement and routing variables encoded in QUBO form. 

\end{itemize}

\subsubsection{Wireless physical-layer applications}\label{sec:wireless_apps}

\begin{itemize}

\item \emph{Beamforming and phase-shift optimization}: In multiple-input multiple-output (MIMO) beamforming design, and in the optimization of reconfigurable surfaces' phase-shifters, the transmitted, received, or reflected symbols are constrained to discrete phase or constellation states, allowing the corresponding NP combinatorial optimization problem to be solved by quantum computing means.

\item \emph{Channel decoding}: In channel decoding, the binary codeword variables can be likewise embedded into an Ising optimization framework. For polar codes, decoding can be formulated as a QUBO over the encoding graph with node and frozen-bit constraints. For LDPC codes, the cost Hamiltonian combines parity check penalties with channel reliability terms. so that low energy states of the Hamiltonian correspond to valid codewords.

\end{itemize}
\subsection{Quantum Machine Learning for Communications}\label{sec:qml}

Quantum machine learning constitutes the second major computational pattern relevant to QB. In contrast to quantum optimization, where the objective is to solve a reduced combinatorial problem, QML is concerned with learning, prediction, detection, and decision-making from data~\cite{chehimiQFL}. In the NISQ era, the most practical realization is typically hybrid, with a parameterized quantum circuit~(PQC) used along with classical learning and trained by a classical optimizer \cite{Zhang2025HybridQAE}. 

Some representative examples are given below.
\begin{itemize}
\item \emph{Quantum kernel methods for backbone anomaly detection}: Backbone anomaly detection can be formulated as a learning task over high-dimensional telemetry, including flow statistics, performance metrics, and security events collected across PoPs. Quantum models can then encode these feature vectors into quantum states and use PQCs to extract compressed representations, using methods such as quantum  support vector machines and quantum $k$-nearest-neighbor classifiers.
\item \emph{Hybrid quantum-classical autoencoders for end-to-end communication}: The transmitter employs PQCs to learn the encoded signal representation while the receiver remains classical. Source messages are embedded into quantum states, processed through layered PQCs and then transmitted over the channel. The receiver performs decoding through standard neural-network layers, enabling joint transmitter-receiver optimization.

\item \emph{Quantum reinforcement learning for wireless resource allocation}: PQCs or quantum neural networks are embedded within the policy or value-function approximation loop to address tasks such as power allocation, user scheduling, and channel assignment. The wireless environment is modeled as a Markov decision process, where then, quantum-enhanced agents encode system states into quantum representations.

\item \emph{Quantum federated learning~(QFL) across PoPs}: Distributed PoPs collaboratively train a shared QML model, by locally training a quantum model on their own data, while only the individual learned PQC parameters of each PoP are periodically exchanged and aggregated through a backbone-coordinated process such as federated averaging. This approach can enable decentralized quantum learning over classical communication infrastructure.

\end{itemize}


\section{Case Study: Deadline-Aware Orchestration of Distributed Quantum Jobs}\label{sec:usecase}

Having mapped QB's runtime to representative communications workloads, we now stress-test its orchestration layer under realistic near-term hardware and deadline constraints. We focus on the distributed-execution mode (Sec.~\ref{sec:exec_modes}), invoked when a workload exceeds the capacity of a single QPU and Layer~2 partitions the reduced circuit into subcircuits that must run across the QPU pool. This mode is the most demanding on Layer~3: while the QIP still decides whether to admit each job onto the quantum path, it is the Layer-3 scheduler that must distribute sampling shots across QPUs and respect the LOCC-induced precedence constraints introduced by circuit cutting. We instantiate this scheduler concretely, evaluate its throughput under near-term backbone-cloud conditions, and compare it against both heuristic schedulers and ablations that disable specific quantum-orchestration features.

\vspace{-0cm}
\subsection{Distributed Execution in the QB Runtime}\label{sec:usecase_setup}

We consider a backbone deployment with $M = 5$ heterogeneous QPUs, each co-located with classical compute in a metro-core PoP or regional DC and characterized by a qubit capacity $Q_m$ and a maximum feasible circuit depth $D_m$ that reflects gate-error-limited fidelity. Near-term quantum networks do not yet provide reliable inter-PoP entanglement~\cite{chehimi2022physics}, so inter-QPU coordination is restricted to LOCCs. This restriction rules out entanglement-assisted distributed execution and makes \emph{circuit cutting} the relevant Layer-2 technique: a large circuit is partitioned into smaller subcircuits that execute on separate QPUs, and the original circuit's output statistics are reconstructed by classical post-processing. Recovering the accuracy of the uncut circuit, however, requires additional shots-the \emph{sampling overhead}-and, depending on the cut type, may also impose precedence constraints.

Network-intelligence applications (Layer~4) submit jobs to the QB request buffer; each job carries a deadline $\tau_i$ inherited from the control-loop epoch of the originating application. Following the reduce-solve-verify pipeline (Sec.~\ref{sec:qip}), Layer~2 compiles each submitted circuit using a combination of gate cuts~\cite{mitarai2021gatecut} and the shot-efficient LOCC wire-cut technique of~\cite{harada2024doubly}. Per job, the gate cut produces $12$ independent subcircuits with no precedence constraints, while the LOCC wire cut produces $6$ subcircuits ($3$ measurement, $3$ preparation), where each preparation subcircuit requires the classical outcome of its paired measurement subcircuit before it can execute. The compiler therefore emits, for each job $i$, a set of generated subcircuits $\mathcal{S}_i$ together with a precedence graph $\mathcal{P}_i$, represented as a directed acyclic graph~(DAG), that encodes the execution dependencies introduced by the LOCC wire cuts.

The Layer-3 scheduler receives $(\mathcal{S}_i, \mathcal{P}_i)$, along with per-subcircuit qubit demand $Q_{i,j}$, depth $D_{i,j}$, and required shot count $N_{i,j}$. It must make four interlocking decisions: (i)~which jobs to admit, since some may be dropped if their deadlines cannot be met under current load; (ii)~which QPUs are eligible to execute each subcircuit, given $Q_m$ and $D_m$; (iii)~how each subcircuit's sampling shots are split into execution fragments and distributed across eligible QPUs to maximize parallelism; and (iv)~the start time of each fragment, so that LOCC precedence and single-fragment-per-QPU exclusivity are jointly respected. The resulting problem is precedence-constrained, resource-heterogeneous, and admits shot splitting-computationally hard at scale. We solve it with a simulated-annealing metaheuristic that generates feasible neighbours by admitting a previously dropped job, swapping an admitted job for a dropped one, or performing local moves on shot splits and start times, accepting moves under a Metropolis rule with an exponential cooling schedule \cite{dehaini2025deadline}. 


\subsection{Results}\label{sec:usecase_results}

We compare the QB Layer-3 scheduler (\emph{QB-scheduler}) against six baselines spanning both classical heuristics and quantum-cloud-specific schedulers. The heuristic baselines are \emph{greedy} (earliest-deadline-first with shot splitting), \emph{list scheduling} (a classical precedence-respecting list scheduler)~\cite{ferrari_works_2024}, and \emph{random} scheduling. The quantum-cloud baselines are ablations of the QB-scheduler's distinguishing features. \emph{QIP w/o shot distribution}, the shot-agnostic scheduler of~\cite{bhoumik2023notads}, treats each subcircuit's shots as a single indivisible job. \emph{QIP w/o LOCC awareness} replaces the LOCC wire cut of~\cite{harada2024doubly} with the dependency-free wire cut of~\cite{peng2020simulating}, avoiding precedence at the cost of $16$ instead of $6$ subcircuits and a higher sampling overhead. \emph{QIP w/o either} disables both features. All baselines are made deadline-aware for a fair comparison.

\begin{figure}[t]
\centering
\includegraphics[width=\columnwidth]{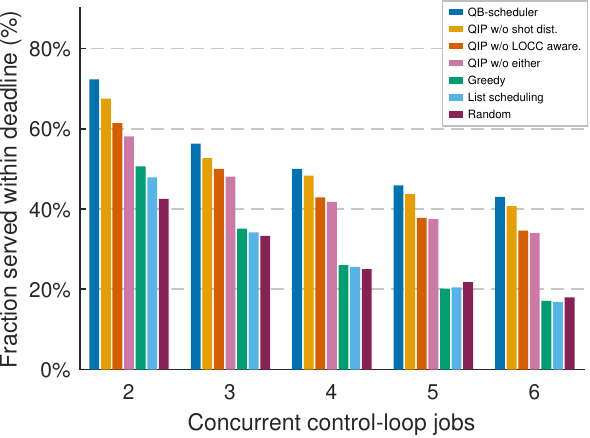}
\caption{Average fraction of jobs completed within deadline vs.\ number of concurrent control-loop jobs. The QB-scheduler consistently outperforms both heuristic and quantum-cloud-specific baselines; the gap widens with load.}
\label{fig:served_vs_jobs}
\vspace{-0cm}
\end{figure}

\begin{figure}[t]
  \centering
  \begin{subfigure}[t]{\columnwidth}
    \centering
    \includegraphics[width=.7\columnwidth]{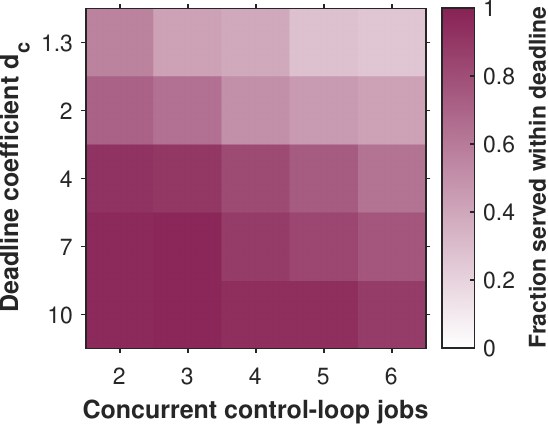}
    \caption{QB-scheduler}
    \label{fig:heatmap_split}
  \end{subfigure}\hfill
  \begin{subfigure}[t]{\columnwidth}
    \centering
    \includegraphics[width=.7\columnwidth]{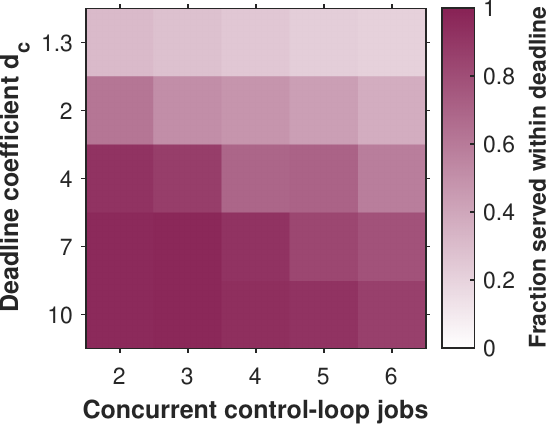}
    \caption{QIP w/o shot distribution}
    \label{fig:heatmap_nosplit}
  \end{subfigure}
  \caption{Fraction of jobs served within deadline vs.\ deadline-tightness coefficient $d_c$ and number of concurrent jobs, for (a) the QB-scheduler, (b) the \emph{QIP w/o shot distribution} ablation.}
  \label{fig:heatmap_dc}
  \vspace{-0cm}
\end{figure}

Fig.~\ref{fig:served_vs_jobs} reports the fraction of jobs served within deadline as the load on the Layer-3 orchestrator increases. The QB-scheduler outperforms the \emph{QIP w/o LOCC awareness} baseline by $8.16\%$ on average, the \emph{QIP w/o either} by $9.60\%$, and \emph{QIP w/o shot distribution} by $2.92\%$ on average, under the default mixed-deadline setup. The heuristic baselines trail by $23.7\%$ (greedy), $24.5\%$ (list), and $25.4\%$ (random). The takeaway is twofold. Generic heuristics, even when made deadline-aware, fail in a near-term distributed-QPU environment because they ignore quantum-specific structure, most notably shot splitting and LOCC-induced precedence. Among quantum-cloud-specific schedulers, the QB-scheduler's joint awareness of shot distribution and LOCC precedence remains measurably valuable, even when the baselines already incorporate sophisticated quantum-cloud heuristics.

Fig.~\ref{fig:heatmap_dc} makes the deadline-regime dependence explicit. When the deadline coefficient $d_c$ is large, i.e., generous deadlines, most schedulers can admit most jobs and the QB-scheduler's advantage over the \emph{QIP w/o shot distribution} baseline is small ($\sim\!1\%$). As the deadline coefficient tightens, the advantage grows. Specifically, we get an advantage of $4.8\%$ for relatively generous deadlines, $6.2\%$ for less generous deadlines, and a $12.8\%$ average advantage for strict deadlines, with a peak of $25\%$ when two concurrent jobs of $d_c=1.3$ are jointly considered. It is noted, that is precisely the regime that 6G backbone control loops will be called to excel, scheduling multiple concurrent jobs with demanding deadlines. 

\vspace{-0cm}
\subsection{Discussion}\label{sec:usecase_discussion} 
The results essentially highlight that distributed quantum execution is not only a hardware-scaling problem, but also a network-orchestration problem. When several quantum jobs compete for heterogeneous QPUs under tight deadlines, the scheduler must manage both the distribution of sampling shots and the  LOCC induced precedence constraints that circuit cutting introduces. Ignoring either aspect turns out to reduce the effective use of QPU capacity, even when the scheduler remains deadline-aware. Thus, the results  reinforce the role of Layer~3 as a load-bearing component of QB, since quantum resources become useful for backbone control only when they are invoked, partitioned and scheduled according to the timing constraints of the network.

\vspace{-0cm}
\section{Open Problems and Future Directions}\label{sec:open_problems}

Several challenges remain before QB can transition from an architectural concept to an operational communication platform. These challenges span both the QB runtime itself and the communication workloads executed on top of it.

\vspace{-0cm}
\subsection{Architectural challenges}

\subsubsection{Scalable orchestration} The Layer-3 scheduler studied in Sec.~\ref{sec:usecase} operates at the scale of a few distributed QPUs and concurrent jobs. Extending QB to large backbone deployments with many PoPs and continuously arriving workloads will likely require distributed or learning-based orchestration methods capable of preserving deadline- and dependency-awareness without the overhead of centralized scheduling.

\subsubsection{Control-loop latency} QPU compilation, queueing, execution, and classical post-processing must all fit within the timing constraints of network control loops. Techniques such as pre-compiled optimization templates, rolling-horizon scheduling, and workload batching may reduce orchestration overhead, but systematic benchmarking against realistic communication control-plane deadlines remains unexplored.

\subsubsection{Interoperability and standardization}
No standardized interface currently exists for invoking quantum services from communication controllers. Interoperability between QB primitives and emerging orchestration frameworks, such as ETSI NFV MANO and O-RAN remains an open challenge.

\subsubsection{Energy and operational cost} Any claimed quantum advantage must be evaluated holistically, including cryogenic cooling overhead, classical orchestration cost, and GPU-accelerated baselines. Practical deployment therefore requires system-level metrics that quantify latency, energy consumption, and operational efficiency on a per-decision basis.


\vspace{-0.2cm}
\subsection{Application-level challenges}

\subsubsection{Problem reduction} Most backbone and wireless optimization problems remain far beyond the scale of near-term QPUs. The effectiveness of QB therefore depends heavily on classical preprocessing steps such as topology reduction, traffic aggregation, and constraint verification, which determine if the reduced quantum workload is operationally useful.

\subsubsection{Solution reliability} Near-term QPU outputs are inherently noisy and probabilistic. Practical deployment will therefore require confidence estimation, classical verification, and fallback mechanisms before optimization decisions can be safely applied to operational networks.

\subsubsection{QML under shared runtime constraints} In QB, QML workloads share the same QPU infrastructure as latency-sensitive optimization tasks. This raises operational questions related to shot allocation, calibration and drift control, and workload prioritization, that extend beyond the design of the QML algorithms themselves.

\section{Conclusion}\label{sec:conclusion}

Quantum computing is widely envisioned as a potential accelerator for future communication networks, yet its practical integration into existing communication infrastructure requires more than just access to quantum hardware. This article proposed QB, a hybrid quantum-classical control plane for communication and network control, and provided a systematic framework through which quantum optimization and QML workloads can be integrated into backbone and wireless control loops. A preliminary case studied showed that QB can significantly improve workload execution under tight deadline constraints, highlighting its potential role in future communication infrastructures. However, fully achieving the QB vision will require interdisciplinary progress in architectural and application-level challenges, all of which define a concrete research roadmap for quantum-enabled network intelligence.

\bibliographystyle{IEEEtran}
\bibliography{References}

@article{fan2022hybrid,
  title={Hybrid quantum-classical computing for future network optimization},
  author={Fan, Lei and Han, Zhu},
  journal={IEEE Netw.},
  volume={36},
  number={5},
  pages={72--76},
  year={2022},
  publisher={IEEE}
}

@inproceedings{dehaini2025deadline,
  title={Deadline-Aware Scheduling of Distributed Quantum Circuits in Near-Term Quantum Cloud},
  author={Dehaini, Nour and Chahoud, Christia and Chehimi, Mahdi},
  booktitle={Proc. 2026 IEEE Int. Conf. Quantum Commun., Netw., Comput. (QCNC)},
  pages={622--629},
  year={2026},
  organization={IEEE}
}

@article{peng2020simulating,
  title={Simulating large quantum circuits on a small quantum computer},
  author={Peng, Tianyi and Harrow, Aram W and Ozols, Maris and Wu, Xiaodi},
  journal={Phys. Rev. Lett.},
  volume={125},
  number={15},
  pages={150504},
  year={2020},
  publisher={APS}
}

@article{carrera2024combining,
  title={Combining quantum processors with real-time classical communication},
  author={Carrera Vazquez, Almudena and Tornow, Caroline and Rist{\`e}, Diego and Woerner, Stefan and Takita, Maika and Egger, Daniel J},
  journal={Nature},
  volume={636},
  number={8041},
  pages={75--79},
  year={2024},
  publisher={Nature Publishing Group UK London}
}

@article{chehimi2022physics,
  title={Physics-informed quantum communication networks: A vision toward the quantum internet},
  author={Chehimi, Mahdi and Saad, Walid},
  journal={IEEE Netw.},
  volume={36},
  number={5},
  pages={32--38},
  year={2022},
  publisher={IEEE}
}

@article{Zhang2025HybridQAE,
  author  = {Bolun Zhang and Gan Zheng and Nguyen Van Huynh},
  title   = {A Hybrid Quantum-Classical Autoencoder Framework for End-to-End Communication Systems},
  journal = {IEEE Wirel. Commun. Lett.},
  volume  = {14},
  number  = {3},
  pages   = {806--810},
  year    = {2025},
  month   = mar,
  doi     = {10.1109/LWC.2024.3524330}
}

@ARTICLE{Krikidisris,
  author={Krikidis, Ioannis and Gilbert, Valentin},
  journal={IEEE Commun. Mag.}, 
  title={{Quantum Optimization in Wireless Communication Systems: Principles and Applications}}, 
  year={2026},
  volume={},
  number={},
  pages={1-7},
  doi={10.1109/MCOM.001.2500231}}

@INPROCEEDINGS{chehimiQFL,
  author={Chehimi, Mahdi and Saad, Walid},
  booktitle={Proc. IEEE Int. Conf. Acoust., Speech Signal Process. (ICASSP), 2022}, 
  title={Quantum Federated Learning with Quantum Data}, 
  year={2022},
  volume={},
  number={},
  pages={8617-8621},
  doi={10.1109/ICASSP43922.2022.9746622}}

@article{caleffi2024dqc,
  author={Caleffi, M. and Amoretti, M. and Ferrari, D. and Illiano, J.
  and Manzalini, A. and Cacciapuoti, A.~S.},
  title={Distributed quantum computing: a survey},
  journal={Computer Networks}, volume={254}, pages={110672}, year={2024}}

@article{barral2025dqc,
  author={Barral, D. and others},
  title={Review of distributed quantum computing: From single QPU to
  high performance quantum computing},
  journal={Computer Science Review}, volume={57}, pages={100747}, year={2025}}

@article{mitarai2021gatecut,
  author={Mitarai, K. and Fujii, K.},
  title={Constructing a virtual two-qubit gate by sampling single-qubit operations},
  journal={New Journal of Physics}, volume={23}, number={2}, pages={023021}, year={2021}}

@article{harada2024doubly,
  author={Harada, H. and others},
  title={Doubly optimal parallel wire cutting without ancilla qubits},
  journal={PRX Quantum}, volume={5}, number={4}, pages={040308}, year={2024}}

@inproceedings{ferrari_works_2024,
  author={Ferrari, D. and Amoretti, M.},
  title={A design framework for the simulation of distributed quantum computing},
  booktitle={Workshop on HPC and Quantum Computing Integration},
  pages={4--10}, year={2024}}

@article{bhoumik2023notads,
  title={Distributed scheduling of quantum circuits with noise and time optimization},
  author={Bhoumik, Debasmita and Majumdar, Ritajit and Saha, Amit and Sur-Kolay, Susmita},
  journal={arXiv preprint arXiv:2309.06005},
  year={2023}
}

@ARTICLE{9390169,
  author={Tataria, Harsh and Shafi, Mansoor and Molisch, Andreas F. and Dohler, Mischa and Sjöland, Henrik and Tufvesson, Fredrik},
  journal={Proc. IEEE}, 
  title={{6G Wireless Systems: Vision, Requirements, Challenges, Insights, and Opportunities}}, 
  year={2021},
  volume={109},
  number={7},
  pages={1166-1199},
  doi={10.1109/JPROC.2021.3061701}}
\vspace{-1cm}
\begin{IEEEbiographynophoto}{
Mahdi Chehimi} (Member, IEEE) is an Assistant Professor of Electrical and Computer Engineering at the American University of Beirut (AUB). He received his Ph.D. in Electrical and Computer Engineering from Virginia Tech in 2024. His research interests include quantum communication networks, distributed quantum computing, sensing and machine learning. He organized multiple
quantum workshops and tutorials at flagship IEEE conferences, and received the best paper award in IEEE QCE 2023 and QCNC 2026.
\end{IEEEbiographynophoto}
\vspace{-.8cm}
\begin{IEEEbiographynophoto}{Nour Dehaini} is a Graduate Research Assistant in the Department of Electrical and Computer Engineering at the American University of Beirut. Her research interests include distributed quantum computing, and she received the IEEE QCNC 2026 best paper award.
\end{IEEEbiographynophoto}
\vspace{-.8cm}
\begin{IEEEbiographynophoto}{Nikos A. Mitsiou} (Member, IEEE) is a Post-Doctoral Researcher in the Department of Electrical and Computer Engineering, University of Cyprus, Nicosia. He received the IEEE WCNC 2025 Best Paper Award. His research interests include the intersection of optimization theory and wireless networks.
\end{IEEEbiographynophoto}
\vspace{-.8cm}
\begin{IEEEbiographynophoto}{Ioannis Krikidis}(Fellow, IEEE) (krikidis@ucy.ac.cy) received his Ph.D degree from École Nationale Supérieure des Télécommunications (ENST), Paris, France, in 2005. He is currently a Professor at the Department of Electrical and Computer Engineering, University of Cyprus, Nicosia. His current research interests include wireless communications, 6G, wireless powered communications, RIS, quantum information processing.
\end{IEEEbiographynophoto}
\vspace{-.8cm}
\begin{IEEEbiographynophoto}{Gan Zheng} (Fellow, IEEE) is Professor of connected systems at the School of Engineering, University of Warwick, UK. His research interests include machine learning and quantum computing for wireless communications, reconfigurable intelligent surface, and fluid antennas. He received the 2013 IEEE Signal Processing Letters Best Paper Award, the 2015 GlOBECOM Best Paper Award and the 2018 IEEE Technical Committee on Green Communications and Computing Best Paper Award. He was listed as a Highly Cited Researcher by Thomson Reuters/Clarivate Analytics in 2019. 
\end{IEEEbiographynophoto}

\end{document}